\documentclass[12pt,a4paper]{article}
\usepackage{graphicx}
\begin{document}
 
 \thispagestyle{empty}
 
 \title{Platonic polyhedra tune the 3-sphere:\\ III. Harmonic analysis on  octahedral spherical 3-manifolds.}
 \author{Peter Kramer,\\ Institut f\"ur Theoretische Physik der Universit\"at  T\"ubingen,\\ Germany.}
 \maketitle

\section*{Abstract.}
From the homotopy groups of three distinct octahedral  spherical 3-manifolds we construct 
the isomorphic groups $H$ of deck transformations acting on the 3-sphere.
The $H$-invariant polynomials on the 3-sphere constructed by representation theory span the bases for the harmonic analysis 
on three  spherical manifolds. Analysis of the Cosmic Microwave Background in terms of these new bases can reveal 
a non-simple topology of the space part of space-time.

\section{Introduction.}
We view a spherical topological 3-manifold ${\cal M}$, see \cite{SE34} and \cite{TH97}, as a prototile  on its  
cover $\tilde{{\cal M}}=S^3$. We studied in \cite{KR08} the isometric actions of $O(4,R)$ on the 3-sphere $S^3$ and gave its 
basis as  well-known homogeneous Wigner polynomials in \cite{KR05} eq.(37).
An  algorithm due to Everitt in \cite{EV04} generates  the homotopies for all  spherical 
3-manifolds ${\cal M}$ from five Platonic polyhedra. Using intermediate Coxeter groups, we construct  deck transformations 
acting on the 3-sphere as
isomorphic images \cite{SE34} of homotopies and  generate  
the groups $H={\rm deck}({\cal M}) \sim \pi_1({\cal M})$. 
Following work on the Poincar\'e dodecahedral \cite{KR05}, \cite{KR06}, the tetrahedral \cite{KR08}, and two cubic spherical manifolds \cite{KR09},
we turn here to three octahedral spherical manifolds denoted in \cite{EV04} as $N4, N5, N6$. We construct a basis for the
harmonic analysis on each manifold from $H$-invariant polynomials on the 3-sphere.

One field of applications for  harmonic analysis  is cosmic topology \cite{LA95}, \cite{LU03}: 
The topology of a 3-manifold ${\cal M}$ is favoured if 
data from the Cosmic Microwave Background can be  
expanded in its harmonic basis. The present work provides three novel octahedral 3-manifolds for this analysis.
For the notions of homotopic boundary conditions and random  point symmetry we refer to \cite{KR09b}.

\section{The Coxeter group $G$ and the $24$-cell on $S^3$.}

The cartesian coordinates $x= (x_0,x_1,x_2,x_3)\in E^4$ for $S^3$ we combine   as in 
\cite{KR05}, \cite{KR08} in the matrix form
\begin{equation}
 \label{oc1}
u=
\left[
\begin{array}{ll}
 z_1&z_2\\
-\overline{z}_2&\overline{z}_1\\
\end{array}
\right],\:
z_1=x_0-ix_3,\; z_2=-x_2-ix_1,\: z_1\overline{z}_1+z_2\overline{z}_2=1.
\end{equation}
For the group action we start from the Coxeter group $G<O(4,R)$ \cite{HU90}, \cite{EV04} p. 254,
with the diagram
\begin{equation}
\label{oc2}
G=: \circ\stackrel{3}{-}\circ\stackrel{4}{-}\circ\stackrel{3}{-}\circ.
\end{equation}

\begin{center}
\includegraphics{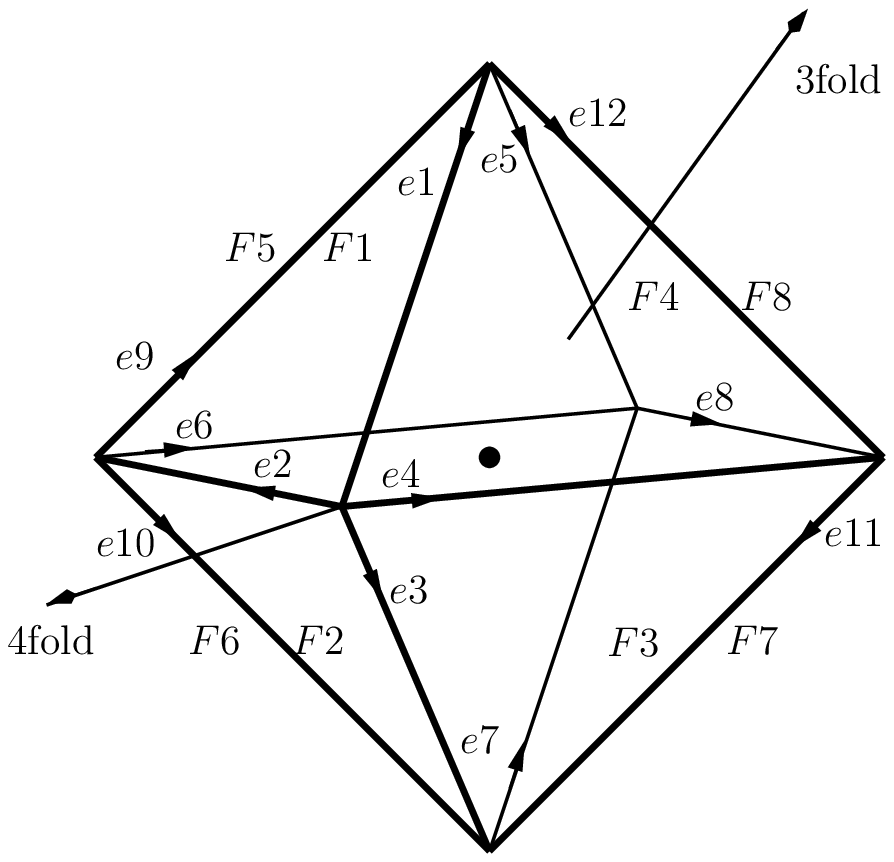}
\end{center}

Fig. 1. The octahedron projected to the plane with faces $F1\ldots F8$ and directed edges $e1\ldots e12$
 according to \cite{EV04}. The products of Weyl reflections $(W_1W_2)$ and $(W_2W_3)$ generate right-handed
$3$fold and $4$fold rotations respectively.

\vspace{0.3cm}

For the Coxeter diagram  eq. \ref{c2a} we  give for the $4$ Weyl reflections 
$W_s, s=1,2,3,4$
the Weyl vectors $a_s$ in {\bf Table 2.1} and compute for each 
$a_s=(a_{s0},a_{s1},a_{s2},a_{s3})$ 
the matrix 
\begin{equation}
\label{c2a}
v_s:=
\left[
\begin{array}{ll}
a_{s0}-ia_{s3}&-a_{s2}-ia_{s1}\\
a_{s2}-ia_{s1}&a_{s0}+ia_{s3}
\end{array} \right] \in SU(2,C).
\end{equation}

The matrices $v_s$  are used  to relate, see  \cite{KR08}, the 
Weyl reflections to  $(SU^l(2,C)\times SU^r(2,C))$ acting by left and right multiplication 
on the coordinates eq. \ref{oc1}. 
We include  the (orientation preserving) inversion $J_4\in G$,  
and list the additional Weyl reflection $W_0$.
The Coxeter group eq. \ref{oc2} is of order $|G|=48 \times 24= 1152$.
The first three Weyl reflections from {\bf Table 2.1} generate, see \cite{HU90},  
the  cubic Coxeter subgroup 
\begin{equation}
 \label{oc4}
O=: \circ\stackrel{4}{-}\circ\stackrel{3}{-}\circ,
\end{equation}
isomorphic to the octahedral group $O \sim (C_2)^3 \times_s S(3)$
acting on $E^3\in E^4$. 
The octahedral tiling of $S^3$ is the 24-cell discussed in \cite{SO58} pp. 171-2. The center positions of the 24 octahedra in the octahedral 24-cell tiling are the midpoints of the 24 square faces of the 8 cubes in the 8-cell tiling shown in \cite{KR09},
Fig. 1. As shown in \cite{SO58} pp. 178-9,  vertices of six octahedra are  located  at each  center of a cube from  the 8-cell.

{\bf Table 2.1}
The Weyl vectors $a_s,\: s=1,.., 4$ and $a_0$ for the Coxeter group $G$ eq. \ref{oc2}, and the
$2 \times 2$ unitary matrices $v_s$ eq. \ref{c2a}, in terms of $\theta: = \exp(i\pi/4)$. 
\begin{eqnarray*}
\label{oc6}
&&\begin{array}{|l|l|l|}
\hline
s &{\rm Weyl}\:{\rm vector}\:a_s &{\rm matrix}\: v_s\\
\cline{1-3}
&&\\
1 & (0,\sqrt{\frac{1}{2}},-\sqrt{\frac{1}{2}},0)
&\left[
\begin{array}{ll}
0&\overline{\theta}\\
-\theta&0\\
\end{array}
\right]\\
2 & (0,0,-\sqrt{\frac{1}{2}},\sqrt{\frac{1}{2}})
&\sqrt{\frac{1}{2}}
\left[\begin{array}{ll}
-i&1\\
-1&i\\
\end{array}
\right]\\
3 & (0,0,0,1)
&
\left[
\begin{array}{ll}
-i&0\\
0&i\\
\end{array}
\right]\\
4 & (\frac{1}{2},\frac{1}{2},\frac{1}{2},\frac{1}{2})
&\sqrt{\frac{1}{2}}\left[
\begin{array}{ll}
\overline{\theta}&-\theta\\
\overline{\theta}&\theta\\
\end{array}
\right]\\
0 & (1,0,0,0)
&\left[
\begin{array}{ll}
1&0\\
0&1\\
\end{array}
\right]\\
\hline
\end{array}
\end{eqnarray*}

\section{From  homotopies to deck transformations.}

\subsection{Generators.}
The spherical Coxeter group $G$ is generated by the Weyl reflections $W_s$ given in {\bf Table 2.1}. 
In the next section we give for the three octahedral manifolds $N4, N5, N6$ the edge gluing schemes computed in 
\cite{EV04}, but  include  the  corrections given in  \cite{CAV09}. These corrections apply in particular to the manifold $N5$.  
The construction proceeds in the following steps:\\
(i) An  edge gluing scheme  lists glued triples of oriented edges for pairs of glued faces $Fi\cup Fj$ in its rows. 
The four generators of
the first homotopy group  each prescribe  a gluing  of three oriented chains of edges, bounding  counterclockwise a preimage face Fi
and clockwise an image face Fj of the prototile. These chains taken from Fig. 1 are given between square brackets.\\ 
(ii) Any deck transformation is constructed from a homotopy by first rotating the preimage face Fi wrt. the center $(1,0,0,0)$ of the 
prototile  to the position of face F4, and then applying a rotation $(W_1W_2)^{\nu},\: \nu=0,1,2$ 
preserving the center of F4. Inversion $J_3$ in the center  of the prototile then maps the preimage face from the position of face F4 to the one of F6.  This inversion can be expressed  as $J_3=J_4 W_0$. 
The total inversion $J_4$ preserves orientation and commutes with all rotations. Applying the Weyl reflection $W_4$, the preimage face now in position F6 is mapped into itself, while the octahedral prototile  is mapped into an image tile. 
\\ 
(iii) By a final  rotation wrt. the center of the prototile, the preimage face is mapped from the position of F6 into 
the image position of face Fj. An  appropriate choice of $\nu$ yields the  edge mapping prescribed by the homotopy. 
By virtue of the Weyl reflection, the image face Fj 
separates the prototile from a fixed octahedral image. The orientation of the
chain of edges of the image face now is counterclockwise when referred to the center of the image tile. The map from the prototile 
to the image tile in this position is the deck transformation isomorphic to the homotopy.
 
All the  operations in (i-iii) are elements of the Coxeter group $G$ and moreover of $SO(4,R)$.
The rotations are generated from the 3fold rotation $(W_1W_2)$ and the 4fold
rotation $(W_2W_3)$, indicated  in Fig. 1. 
Any Weyl reflection  $W_s$ is associated with a $2 \times 2$ matrix $v_s$ given in {\bf Table 2.1}.
Products of two Weyl reflections generate 
rotations. The  conversion from these products 
to rotations  $g= (w_l,w_r)$ is given from \cite{KR08} eq. (60) by 
\begin{equation}
\label{oc7}
(W_iW_j) \rightarrow T_g= T_{(w_l,w_r)},\: g=(w_l,w_r)=(v_iv_j^{-1}, v_i^{-1}v_j).
\end{equation}
The operator $T_g$ acts on functions $f(u)$ on $S^3$ in coordinates $u$  from eq. \ref{oc1} as
\begin{equation}
\label{oc8}
(T_{(w_l,w_r)}f)(u):= f(w_l^{-1}uw_r). 
\end{equation} 

Any  product of the in general five operations described under (i-iii) is a deck transformation,  preserves  orientation, and is isomorphic to 
a homotopic gluing. We list them  for the three manifolds. Finally the deck transformations are converted  by use of eq. \ref{oc7}, {\bf Table 2.1},  and multiplication into  pairs $(w_l,w_r) \in (SU^r(2,C)  \times SU^r(2,C))$, given in the following Tables.

The (isomorphic) groups $H$ of homotopies and of deck transformations, distinct for different manifolds, all have order $24$ equal to the number
of octahedral tiles. 
These groups if not abelian must appear in the  Table of Coxeter and Moser \cite{CO65} pp. 134-5.

\subsection{Center positions under deck transformations.}

From the coordinates eq. \ref{oc1} of $S^3$, the center $u=e$ of the octahedral prototile is transformed by $g=(w_l,w_r)$ into an image  center 
\begin{equation}
\label{oc8a}
 g=(w_l, w_r)\in H: e \rightarrow u'=w_l^{-1}w_r.
\end{equation}

Since all three groups of deck transformations must produce the same 24-cell, it follows that their lists of octahedral centers must coincide
up to permutations. For the manifold $N6$ we shall find that its group of deck transformations is the binary tetrahedral group 
${\cal T}$ with all elements of the form $g=(w_l,e)$. From eq. \ref{oc8a} it then follows that the list of the 24 octahedral centers $u'$ in the 24-cell
can be written as 
\begin{equation}
\label{oc8b}
N6: g=(w_l,e)\in {\cal T},\: u'=w_l^{-1}.
\end{equation}
with $w_l$ running over all group elements in {\bf Table 6.2}. The elements $g=(w_l,w_r)$ of the groups $H\neq {\cal T}$ for the manifolds $N4, N5$ therefore must reproduce 
by the products $u'$ in eq. \ref{oc8a} these $24$ center positions. For the manifold $N5$ we display this relation in {\bf Table 5.2}.

\subsection{Harmonic analysis on octahedral 3-manifolds.}
Once we have derived the explicit matrix form of the three groups $H$ of deck transformations, we have all the tools
for the harmonic analysis. From any $g=(w_l,w_r)\in SO(4,R)$ we can, as outlined in general in \cite{KR08} eq.(44), pass to its representations $D^{(j,j)}(g)= D^j(w_l) \times D^j(w_r)$
by use of Wigner representation matrices $D^j$ of $SU(2,C)$.
From these representations we can construct the general projection and Young operators 
\cite{KR08} eq. (82) to $H$-invariant polynomials of fixed $j$ and degree $2j$. The projection yields linear combinations of spherical harmonics
or Wigner polynomials  $D^j_{m_1,m_2}(u)$ of degree $2j$. For the octahedral manifold $N4$ we give  the final result of this projection in {\bf Table 4.3}.
The characters follow from \cite{KR08} eq. (45) and allow to derive by \cite{KR08} eq.(62) the
multiplicities for any degree $2j$ and group $H$.

\section{Manifold N4}

Face gluings:
\begin{equation}
 \label{oc8c}
F6\cup F2,\: F5\cup F3,\: F1\cup F4,\: F7\cup F8.
\end{equation}

Edge gluing scheme:
\begin{equation}
\label{oc9}
\left[
\begin{array}{lll}
1&4&9\\
2&7&\overline{12}\\
3&6&\overline{10}\\
5&8&11\\
\end{array}
\right]
\end{equation}
Edge and face gluing generators of $\pi_1(N4)$:
\begin{eqnarray}
\label{oc10}
&&g_1: 6\cup 2,\:
 \left[
\begin{array}{lll}
\overline{10}&&\overline{7}\\
&6&
\end{array}
\right]
\rightarrow
\left[
\begin{array}{lll}
3&&\overline{2}\\
&\overline{10}&
\end{array}
\right],\:
\\ \nonumber
&&g_2: 5 \cup 3,\: 
\left[
\begin{array}{lll}
5&&9\\
&\overline{6}&
\end{array}
\right]
\rightarrow
\left[
\begin{array}{lll}
11&&4\\
&\overline{3}&
\end{array}
\right],\:
\\ \nonumber
&&g_3: 1 \cup 4,\: 
\left[
\begin{array}{lll}
\overline{9}&&\overline{1}\\
&\overline{2}&
\end{array}
\right]
\rightarrow
\left[
\begin{array}{lll}
\overline{1}&&\overline{4}\\
&12&
\end{array}
\right],\:
\\ \nonumber
&&g_4: 7\cup 8,\:
\left[
\begin{array}{lll}
7&&11\\
&8&
\end{array}
\right]
\rightarrow
\left[
\begin{array}{lll}
\overline{12}&&8\\
&5&
\end{array}
\right],\:
\end{eqnarray}
Isomorphic generators of $deck(N4)$:
\begin{eqnarray}
\label{oc11}
&&g_1=(W_2W_3)^2(W_1W_2)(W_4W_0) J_4 
\\ \nonumber
&&g_2=(W_3W_2)(W_1W_2)(W_4W_0)(W_2W_3)J_4
\\ \nonumber
&&g_3=(W_2W_1)(W_4W_0)(W_2W_3)^2(W_2W_1)J_4
\\ \nonumber
&&g_4=(W_1W_2)(W_3W_2)(W_2W_1)(W_4W_0)(W_3W_2)J_4
\end{eqnarray}

{\bf Table 4.1} Generators $g=(w_l,w_r)$ of $deck(N4)$ in the scheme eqs. \ref{oc7},\ref{oc8}.
We use the short-hand notation of {\bf Table 6.2}.
\begin{eqnarray*}
\label{oc11b}
&&\begin{array}{|l|l|l|}
\hline
 g& w_l& w_r\\ 
\cline{1-3}
g_1 & -\alpha_2&\mu\\
\hline
g_2 & -\alpha_2^{-1}&-e\\
\hline 
g_3 &\alpha_2&\nu\\
\hline
g_4 &\alpha_2^{-1}&\omega\\ 
\hline
\end{array}
\end{eqnarray*}
The generators $g=(w_l, w_r)$ have for $w_l$ the order $6$ or $3$ and $w_l^3=\pm e$,  for 
$w_r$ the order $4$ and $w_r^2=-e$ . From this it follows that $g_q^3=(\pm e,-w_r)\sim (e, \pm w_r)$. It is easy to see that 
the four elements $g_q^3$ generate the quaternion group by right action which we denote by $Q^r$. Similarly the  powers $4$ of the generators fulfill $g_q^4=(w_l^4, -e)\sim (-w_l^4, e)$ and so act from the left. Inspection of these elements 
shows that they can be written as powers of $(-\alpha_2,e)$. The group generated by them is a cyclic group of order $3$ 
which we denote as $C_3^l$. Now it is easy to conclude that the two subgroups generate the direct product group 
$C_3^l \times Q^r$ of order $24$, compare  Coxeter and Moser \cite{CO65} pp. 134-5, as the group of homotopies and of deck transformations for the 3-manifold $N4$.

{\bf Table 4.2}  The elements $g=(w_l, w_r)$ of the group ${\rm deck}(N4)= C_3^l\times Q^r$ in the notation of {\bf Table 6.2}.
\begin{eqnarray*}
\label{oc11c}
\begin{array}{|l|l|} \hline 
{\rm subgroup}& {\rm elements}\\ \hline
C_3^l& ( -\alpha_2,e),((\alpha_2)^2,e),( (-\alpha_2)^3,e)=(e,e) \\ \hline
Q^r& (e,\pm e), (e,\pm \mu), (e, \pm \nu), (e, \pm\omega)\\ \hline
\end{array} 
\end{eqnarray*}
The 24 center positions  $u'=(w_l^{-1}w_r)$ of $C_3^l\times Q^r$  reproduce the elements of the binary tetrahedral group {\bf Table 6.2}.

For the projection to a $H$-invariant basis we  first diagonalize the generator $-\alpha_2\in C_3^l$,
\begin{eqnarray}
\label{oc11d}
&& -\alpha_2= c 
\left[
\begin{array}{ll}
 \exp(\frac{2\pi i}{3})&0\\
0&\exp(-\frac{2\pi i}{3})
\end{array} \right]
c^{\dagger},
\\ \nonumber
&& c=
\left[
\begin{array}{ll}
(1-i)\frac{-1+\sqrt{3}}{2\sqrt{3-\sqrt{3}}}& -(1-i)\frac{1+\sqrt{3}}{2\sqrt{3+\sqrt{3}}}\\
\frac{1}{\sqrt{3-\sqrt{3}}}& \frac{1}{\sqrt{3+\sqrt{3}}}
\end{array} 
\right].
\end{eqnarray}
Upon the coordinate transform from  $u$ to $c^{\dagger}u$ we can replace the matrix $-\alpha_2$ by its diagonal representative. Now the projection 
to the identity representations of $C^l_3$ simply requires $m_1\rightarrow \rho \equiv 0\:  {\rm mod}\:  3$ and excludes any other value of $m_1$. Next we consider the group $Q^r$ acting from the right. We simply transcribe the result on the group $Q$ from \cite{KR09}, {\bf Table 10} from left to right action.
Combining left and right action into $C^l_3 \times Q^r$ we arrive at the $H$-invariantbasis of the harmonic analysis on $N4$ given in {\bf Table 4.3}.

{\bf Table 4.3}: The $(C_3^l \times Q^r)$-invariantbasis for the manifold $N4$ in terms of Wigner polynomials $D^j$.
Only integer values of $j$ appear. The coordinate transform $u \rightarrow u'=c^{\dagger}u$ in $D^j(u)$  follows with 
$c$ from eq. \ref{oc11d}. 
\begin{eqnarray*}
\label{oc11e}
\begin{array}{|l|} \hline
j={\rm odd},\: j\geq 3,\: m_2={\rm even},\: 0<m_2 \leq j,\: m_1=\rho \equiv 0\: {\rm mod}\: 3:
\\ \hline
\\ 
 \phi^{j{\rm odd}}_{\rho,m_2}= \left[ D^j_{\rho,m_2}(u')-D^j_{\rho,-m_2}(u')\right],
\\
\\ \hline 
j={\rm even},\: m_2=0,\: m_1=\rho \equiv 0\: {\rm mod}\: 3:
\\  \hline
\\
\phi^{j{\rm even}}_{\rho,0}= D^j_{\rho, 0}(u') 
\\
\\ \hline
j \geq 2,\:   {\rm even},\: 0<m_2\leq j,\: m_2={\rm even},\: m_1=\rho \equiv 0\:  {\rm mod}\: 3:
\\ \hline 
\\
\phi^{j{\rm even}}_{\rho,m_2}= \left[D^j_{\rho, m_2}(u') +D^j_{\rho,-m_2}(u')\right] 
\\
\\ \hline
\end{array} 
\end{eqnarray*}
\vspace{0.2cm}

\section{Manifold N5}

Face gluings:
\begin{equation}
 \label{oc11f}
F6\cup F8,\: F1\cup F4,\: F2\cup F7,\: F3\cup F5.
\end{equation}

Edge gluing scheme:
\begin{equation}
\label{oc12}
\left[
\begin{array}{lll}
1&4&9\\
2&\overline{7}&\overline{12}\\
3&6&8\\
5&\overline{10}&11\\
\end{array}
\right]
\end{equation}

Edge and face gluing generators of $\pi_1(N5)$:
\begin{eqnarray}
\label{oc13}
&&g_1: 6\cup 8,\:
\left[
\begin{array}{lll}
\overline{10}&&\overline{7}\\
&6&
\end{array}
\right]
\rightarrow
\left[
\begin{array}{lll}
5&&\overline{12}\\
&8&
\end{array}
\right],\:
\\ \nonumber
&&g_2: 1 \cup 4,\: 
 \left[
\begin{array}{lll}
\overline{9}&&\overline{1}\\
&\overline{2}&
\end{array}
\right]
\rightarrow
\left[
\begin{array}{lll}
\overline{1}&&\overline{4}\\
&12&
\end{array}
\right],\:
\\ \nonumber
&&g_3: 2 \cup 7,\: 
\left[
\begin{array}{lll}
2&&\overline{3}\\
&10&
\end{array}
\right]
\rightarrow
\left[
\begin{array}{lll}
\overline{7}&&\overline{8}\\
&\overline{11}&
\end{array}
\right],\:
\\ \nonumber
&&g_4: 3\cup 5,\:
\left[
\begin{array}{lll}
\overline{11}&&3\\
&\overline{4}&
\end{array}
\right]
\rightarrow
\left[
\begin{array}{lll}
\overline{5}&&6\\
&\overline{9}&
\end{array}
\right],\:
\end{eqnarray}
Isomorphic generators of $deck(N5)$:
\begin{eqnarray}
\label{oc14}
&&g_1=(W_1W_2)(W_3W_2)(W_4W_0) J_4, 
\\ \nonumber
&&g_2=(W_2W_1)(W_4W_0)(W_2W_3)^2(W_2W_1)J_4,
\\ \nonumber
&&g_3=(W_1W_2)(W_2W_3)^2(W_1W_2)(W_4W_0)(W_2W_3)(W_2W_1)J_4
\\ \nonumber
&&g_4=(W_2W_1)(W_2W_3)^2(W_1W_2)(W_4W_0)(W_2W_3)^2(W_1W_2)J_4
\end{eqnarray}

{\bf Table 5.1} Generators $g=(w_l,w_r)$ of $deck(N5)$  with partial use of {\bf Table 6.2}.
Note that the matrices $(w_l,w_r) $ for the generators $g_1, g_3$ do not occur in {\bf Table 6.2} and so do not belong to the 
binary tetrahedral group.
\begin{eqnarray*}
&&\begin{array}{|l|l|l|}
\hline
g& w_l& w_r\\ 
\cline{1-3}
g_1 & 
\sqrt{\frac{1}{2}}\left[
\begin{array}{ll}
-i&-1\\
1&i
\end{array}
 \right]
&
\sqrt{\frac{1}{2}}
\left[
\begin{array}{ll}
 -1&-1\\
1&-1
\end{array}
\right]\\
\cline{1-3}
g_2 & 
\alpha_2&\nu\\
\cline{1-3}
g_3 & 
\left[
\begin{array}{ll}
0&\overline{\theta}\\
-\theta&0
\end{array}
 \right]
&
\sqrt{\frac{1}{2}}
\left[
\begin{array}{ll}
 1&1\\
-1&1
\end{array}
\right]
\\
\cline{1-3}
g_4 & 
\alpha_2&e\\
\hline
\end{array}
\end{eqnarray*}

{\bf Table 5.2} Elements $g_j=(w_l,w_r),\: j= \pm 1,..., \pm 12$ of the group ${\rm deck}(N5)$, enumerated according to the $24$ octahedral center positions $u'=w_l^{-1}w_r\in S^3$,  in the order and notation of  {\bf Table 6.2}.
\begin{eqnarray*}
&&\begin{array}{|l|l|l|l|} \hline
\pm j& w_l         &w_r     &w_l^{-1}w_r\\ \hline
\pm 1&\alpha_2^{-1}& \mp \nu&\pm \alpha_1\\ \hline
\pm 2&\alpha_2^{-1}&\pm e&\pm \alpha_2\\ \hline
\pm 3&\alpha_2     &\pm \nu &\pm \alpha_3\\ \hline
\pm 4&
\sqrt{\frac{1}{2}}
\left[
\begin{array}{ll}
-i&-1\\
1&i
\end{array}
\right]&
\pm \sqrt{\frac{1}{2}}
\left[
\begin{array}{ll}
1&-1\\
1&1
\end{array}
\right]&
\pm \alpha_4\\ \hline
\pm 5&
\sqrt{\frac{1}{2}}
\left[
\begin{array}{ll}
-i&-1\\
1&i
\end{array}
\right]&
\mp \sqrt{\frac{1}{2}}
\left[
\begin{array}{ll}
1&1\\
-1&1
\end{array}
\right]&
\pm \alpha_1^{-1}\\ \hline
\pm 6&\alpha_2&\pm e&\pm \alpha_2^{-1}\\ \hline
\pm 7&
\left[
\begin{array}{ll}
0&\overline{\theta}\\
-\theta&0
\end{array}
\right]&
\mp \sqrt{\frac{1}{2}}
\left[
\begin{array}{ll}
1&-1\\
1&1
\end{array}
\right]&
\pm \alpha_3^{-1}\\ \hline
\pm 8&
\left[
\begin{array}{ll}
0&\overline{\theta}\\
-\theta&0
\end{array}
\right]&
\pm \sqrt{\frac{1}{2}}
\left[
\begin{array}{ll}
1&1\\
-1&1
\end{array}
\right]&
\pm \alpha_4^{-1}\\ \hline
\pm 9&
e&\pm e&\pm e\\ \hline
\pm 10&
-\sqrt{\frac{1}{2}}
\left[
\begin{array}{ll}
i&i\\
i&-i
\end{array}
\right]&
\pm \sqrt{\frac{1}{2}}
\left[
\begin{array}{ll}
1&1\\
-1&1
\end{array}
\right]&
\pm \mu\\ \hline
\pm 11&e&\pm\nu&\pm \nu\\ \hline
\pm 12&
\sqrt{\frac{1}{2}}
\left[
\begin{array}{ll}
i&i\\
i&-i
\end{array}
\right]&
\pm \sqrt{\frac{1}{2}}
\left[
\begin{array}{ll}
1&-1\\
1&1
\end{array}
\right]&
\pm \omega\\ \hline 
\end{array}
\end{eqnarray*}

\section{Manifold N6}
Face gluings:
\begin{equation}
 \label{oc14d}
F6\cup F4,\: F5\cup F3,\: F8\cup F2,\: F7 \cup F1.
\end{equation}

Edge gluing scheme:
\begin{equation}
\label{oc15}
\left[
\begin{array}{lll}
1&8&10\\
2&5&11\\
3&6&12\\
4&7&9\\
\end{array}
\right]
\end{equation}
Edge and face gluing generators of $\pi_1(N6)$:
\begin{eqnarray}
\label{oc16}
&&g_1: 6\cup 4,\:
\left[
\begin{array}{lll}
\overline{10}&&\overline{7}\\
&6&
\end{array}
\right]
\rightarrow
\left[
\begin{array}{lll}
\overline{1}&&\overline{4}\\
&12&
\end{array}
\right],\:
\\ \nonumber
&&g_2: 5 \cup 3,\:
\left[
\begin{array}{lll}
5&&9\\
&\overline{6}&
\end{array}
\right],
\rightarrow
\left[
\begin{array}{lll}
11&&4\\
&\overline{3}&
\end{array}
\right],\:
\\ \nonumber
&&g_3: 8 \cup 2,\:
\left[
\begin{array}{lll}
12&&\overline{5}\\
&\overline{8}&
\end{array}
\right]
\rightarrow
\left[
\begin{array}{lll}
3&&\overline{2}\\
&\overline{10}&
\end{array}
\right],\:
\\ \nonumber
&&g_4: 7\cup 1,\:
\left[
\begin{array}{lll}
7&&11\\
&8&
\end{array}
\right]
\rightarrow
\left[
\begin{array}{lll}
9&&2\\
&1&
\end{array}
\right].
\end{eqnarray}

Isomorphic generators of $deck(N6)$:
\begin{eqnarray}
\label{oc17}
&&g_1=(W_1W_2)(W_4W_0) J_4, 
\\ \nonumber
&&g_2=(W_3W_2)(W_1W_2)(W_4W_0)(W_2W_3)J_4,
\\ \nonumber
&&g_3=(W_2W_3)^2(W_1W_2)(W_4W_0)(W_2W_3)^2J_4,
\\ \nonumber
&&g_4=(W_2W_3)(W_1W_2)(W_4W_0)(W_3W_2)J_4,
\end{eqnarray}

{\bf Table 6.1} Generators $g=(w_l,w_r)$ of $deck(N6)$, compare {\bf Table 6.2}.
\begin{eqnarray*}
&&\begin{array}{|l|l|l|}
\hline 
 g& w_l& w_r\\ 
\cline{1-3}
&&\\
g_1 & 
\sqrt{\frac{1}{2}}\left[
\begin{array}{ll}
\theta&\theta\\
-\overline{\theta}&\overline{\theta}
\end{array}
 \right]:=\alpha_1
&
e\\
\cline{1-3}
&&\\
g_2 & 
\sqrt{\frac{1}{2}}
\left[
\begin{array}{ll}
\overline{\theta}&\theta\\
-\overline{\theta}&\theta
\end{array}
 \right]:=\alpha_2^{-1}
&
e\\
\cline{1-3}
&&\\
g_3 & 
\sqrt{\frac{1}{2}}\left[
\begin{array}{ll}
\overline{\theta}&-\overline{\theta}\\
\theta&\theta
\end{array}
 \right]:= \alpha_4^{-1}
&
e\\
\cline{1-3}
&&\\
g_4 & 
\sqrt{\frac{1}{2}}\left[
\begin{array}{ll}
\theta&-\overline{\theta}\\
\theta&\overline{\theta}
\end{array}
 \right]:=\alpha_3
&
e\\
\hline
\end{array}
\end{eqnarray*}

Using the equivalence $(g_l,g_r)\sim (-g_l, -g_r)$, we can write  $H$ entirely in terms of left actions. The group $H$ of homotopies and deck transformations 
of the 3-manifold $N6$ then turns out to be the
binary tetrahedral group $<2,3,3>$ of order $24$ in the notation of Coxeter and Moser \cite{CO65} pp. 134-5. The elements and multiplication rules are 
given in {\bf Tables 6.2, 6.3}.

{\bf Table 6.2}: The binary tetrahedral group ${\cal T} \sim {\rm deck}(N6)$ has $16$ elements $\pm \alpha_j, \pm \alpha_j^{-1}$ and
$8$ elements $\pm e, \pm \mu, \pm \nu,\pm \omega$, with $\theta=\exp(i\pi/4),\: \overline{\theta}=\exp(-i\pi/4)$.
It acts from the left on $u \in S^3$.
\begin{eqnarray*}
&&\begin{array}{|l|l|l|l|} \hline
\alpha_1&\alpha_2&\alpha_3&\alpha_4\\ 
\cline{1-4}
&&&\\
\sqrt{\frac{1}{2}}
\left[\begin{array}{ll}
\theta &\theta\\
-\overline{\theta}&\overline{\theta}
\end{array}
\right]
&
\sqrt{\frac{1}{2}}
\left[\begin{array}{ll}
\theta &-\theta\\
\overline{\theta}&\overline{\theta}
\end{array}
\right]
& 
\sqrt{\frac{1}{2}}
\left[\begin{array}{ll}
\theta &-\overline{\theta}\\
\theta&\overline{\theta}
\end{array}
\right]
& 
\sqrt{\frac{1}{2}}
\left[\begin{array}{ll}
\theta &\overline{\theta}\\
-\theta&\overline{\theta}
\end{array}
\right]
\\
\cline{1-4}
&&&\\
\alpha_1^{-1}& \alpha_2^{-1}&\alpha_3^{-1}&\alpha_4^{-1}\\
\hline
&&&\\
\sqrt{\frac{1}{2}}
\left[\begin{array}{ll}
\overline{\theta} &-\theta\\
\overline{\theta}&\theta
\end{array}
\right]
& 
\sqrt{\frac{1}{2}}
\left[\begin{array}{ll}
\overline{\theta} &\theta\\
-\overline{\theta}&\theta
\end{array}
\right]
& 
\sqrt{\frac{1}{2}}
\left[\begin{array}{ll}
\overline{\theta} &\overline{\theta}\\
-\theta&\theta
\end{array}
\right]
& 
\sqrt{\frac{1}{2}}
\left[\begin{array}{ll}
\overline{\theta} &-\overline{\theta}\\
\theta&\theta
\end{array}
\right]
\\
\cline{1-4}
&&&\\
e,-e&\mu&\nu& \omega\\
\hline
&&&\\
\left[\begin{array}{ll}
1&0\\
0&1
\end{array}
\right],
-\left[\begin{array}{ll}
1&0\\
0&1
\end{array}
\right]
&
\left[\begin{array}{ll}
0&i\\
i&0
\end{array}
\right]
&
\left[\begin{array}{ll}
0&-1\\
1&0
\end{array}
\right]
&
\left[\begin{array}{ll}
-i&0\\
0&i
\end{array}
\right]
\\
\hline
&&&\\
e^{-1}=e,(-e)^{-1}=-e&\mu^{-1}=-\mu&\nu^{-1}=-\nu& \omega^{-1}=-\omega\\
\hline
\end{array}
\end{eqnarray*}
The elements in this Table obey 
\begin{eqnarray}
&&(\alpha_j)^3=(\alpha_j)^{-3}=-e,\: \frac{1}{2}Tr(\alpha_j)=\frac{1}{2}Tr(\alpha_j^{-1})=
\frac{1}{2},\: j=1,..,4.
\\ \nonumber 
&& \mu^2=\nu^2=\omega^2=-e
\end{eqnarray}
The last four elements generate as subgroup the quaternion group $Q$
of order $8$ with ${\bf i}=-\omega,\:{\bf j}=-\nu,\:{\bf k}=\mu$.

\newpage
{\bf Table 6.3} Multiplication table for $12$ elements $g$ of the binary tetrahedral group ${\rm deck}(N6)$ given in  {\bf Table 6.2}. The $12$ elements $-g$ have been 
suppressed.

\begin{eqnarray*}
&& \begin{array}{|l|llll|llll|llll|} \hline
&\alpha_1&\alpha_2&\alpha_3&\alpha_4&
\alpha_1^{-1}&\alpha_2^{-1}&\alpha_3^{-1}&\alpha_4^{-1}&
\mu&\nu&\omega&e\\ \hline
\alpha_1&
-\alpha_1^{-1}&\alpha_4&-\omega&-\nu&
e&\mu&\alpha_2^{-1}&\alpha_3&
-\alpha_3^{-1}&\alpha_2&\alpha_4^{-1}&\alpha_1\\
\alpha_2&
\alpha_3&-\alpha_2^{-1}&\nu&-\omega&
-\mu&e&\alpha_4&\alpha_1^{-1}&
\alpha_4^{-1}&-\alpha_1&\alpha_3^{-1}&\alpha_2\\
\alpha_3&
\mu&-\omega&-\alpha_3^{-1}&\alpha_1&
\alpha_2&\alpha_4^{-1}&e&\nu&
-\alpha_4&-\alpha_2^{-1}&\alpha_1^{-1}&\alpha_3\\
\alpha_4&
-\omega&-\mu&\alpha_2&-\alpha_4^{-1}&
\alpha_3^{-1}&\alpha_1&-\nu&e&
\alpha_3&\alpha_1^{-1}&\alpha_2^{-1}&\alpha_4\\ \hline
\alpha_1^{-1}&
e&\nu&\alpha_4^{-1}&\alpha_2&
-\alpha_1&\alpha_3^{-1}&-\mu&\omega&
\alpha_2^{-1}&-\alpha_4&-\alpha_3&\alpha_1^{-1}\\
\alpha_2^{-1}&
-\nu&e&\alpha_1&\alpha_3^{-1}&
\alpha_4^{-1}&-\alpha_2&\omega&\mu&
-\alpha_1^{-1}&\alpha_3&-\alpha_4&\alpha_2^{-1}\\
\alpha_3^{-1}&
\alpha_4&\alpha_1^{-1}&e&-\mu&
\omega&-\nu&-\alpha_3&\alpha_2^{-1}&
\alpha_1&\alpha_4^{-1}&-\alpha_2&\alpha_3^{-1}\\
\alpha_4^{-1}&
\alpha_2^{-1}&\alpha_3&\mu&e&
\nu&\omega&\alpha_1^{-1}&-\alpha_4&
-\alpha_2&-\alpha_3^{-1}&-\alpha_1&\alpha_4^{-1}\\ \hline
\mu&
-\alpha_2&\alpha_1&-\alpha_1^{-1}&\alpha_2^{-1}&
\alpha_3&-\alpha_4&\alpha_4^{-1}&-\alpha_3^{-1}&
-e&-\omega&\nu&\mu\\
\nu&
\alpha_4^{-1}&-\alpha_3^{-1}&-\alpha_4&\alpha_3&
-\alpha_2^{-1}&\alpha_1^{-1}&\alpha_2&-\alpha_1&
\omega&-e&-\mu&\nu\\
\omega&
\alpha_3^{-1}&\alpha_4^{-1}&\alpha_2^{-1}&\alpha_1^{-1}&
-\alpha_4&-\alpha_3&-\alpha_1&-\alpha_2&
-\nu&\mu&-e&\omega\\
e&\alpha_1&\alpha_2&\alpha_3&\alpha_4&
\alpha_1^{-1}&\alpha_2^{-1}&\alpha_3^{-1}&\alpha_4^{-1}&
\mu&\nu&\omega&e\\ \hline
\end{array}
\end{eqnarray*}

\section{Conclusion.}
In the present work we extend  
the study of the harmonic analysis on Platonic 3-manifolds beyond  the dodecahedral, the 
tetrahedral and the two cubic spherical manifolds. From homotopy we  construct and identify  three groups $H,\:  |H|=24$ of deck transformations for three octahedral spherical 3-manifolds and 
give their action on the 3-sphere. Representation theory of $SO(4,R)>H$ provides  the tools for the multiplicity and projection of 
$H$-invariant polynomial bases of the harmonic analysis on 
the octahedral 3-manifolds.

\section*{Acknowledgment.}
The author appreciates the assistance of  T Kramer, Institute for Theoretical Physics, University of Regensburg, Germany, who did the essential 
algebraic computations for the Tables. He is indebted to B Everitt, Dept. of Mathematics, University of York, England, UK, for  
references, in particular \cite{CAV09}.

\end{document}